\documentclass[final]{article} 
\usepackage{spconf,amsmath,graphicx}
\usepackage{multirow}
\usepackage{verbatim}
\usepackage{array}
\usepackage{changes}
\definechangesauthor[name={A A}, color=blue]{aa}
\usepackage{textcomp}
\usepackage{comment}
\usepackage{graphics}

\newcolumntype{L}[1]{>{\raggedright\let\newline\\\arraybackslash\hspace{0pt}}m{#1}}
\newcolumntype{C}[1]{>{\centering\let\newline\\\arraybackslash\hspace{0pt}}m{#1}}

\title{Cross lingual transfer learning for zero-resource domain adaptation}
\name{Alberto Abad$^{1,2}$ \qquad Peter Bell$^{2}$ \qquad Andrea Carmantini$^{2}$ \qquad Steve Renals$^{2}$\thanks{This research is based upon work supported in part by the Office of the Director of National Intelligence (ODNI), Intelligence Advanced Research Projects Activity (IARPA), via Air Force Research Laboratory (AFRL) contract \#FA8650-17-C-9117. The views and conclusions contained herein are those of the authors and should not be interpreted as necessarily representing the official policies, either expressed or implied, of ODNI, IARPA, AFRL or the U.S. Government. The U.S. Government is authorized to reproduce and distribute reprints for governmental purposes notwithstanding any copyright annotation therein.
Alberto Abad was supported by Portuguese national funds through \textit{Funda\c{c}\~{a}o para a Ci\^{e}ncia e a Tecnologia} (FCT) with reference UID/CEC/50021/2013.}}

\address{$^1$INESC-ID / Instituto Superior T\'ecnico, University of Lisbon, Portugal\\
$^2$Centre for Speech Technology Research, University of Edinburgh, UK}

\begin{document}
\ninept
\maketitle
\begin{abstract}
We propose a method for zero-resource domain adaptation of DNN acoustic models, for use in low-resource situations where the only in-language training data available may be poorly matched to the intended target domain.  Our method uses a multi-lingual model in which several DNN layers are shared between languages.  This architecture enables domain adaptation transforms learned for one well-resourced language to be applied to an entirely different low-resource language.  \added[id=aa]{First,} to develop the technique we use English as a well-resourced language and take Spanish to mimic a low-resource language. 
Experiments in domain adaptation between the conversational telephone speech (CTS) domain and broadcast news (BN) domain demonstrate a \replaced[id=aa]{29\%}{26\%} relative WER improvement on Spanish BN test data by using only English adaptation data.  Second, we demonstrate the effectiveness of the method for low-resource languages with a poor match to the 
well-resourced language. Even in this scenario, the proposed method achieves relative WER improvements of 18-27\% by using solely English data for domain adaptation.
\added[id=aa]{Compared to other related approaches based on multi-task and multi-condition training, the proposed method is able to better exploit  well-resource language data for improved acoustic modelling of the low-resource target domain. 
}
\end{abstract}
\section{Introduction}
\label{sec:intro}

In automatic speech recognition (ASR), the problem of building acoustic models that behave robustly in different usage domains is still an open research challenge, despite the emergence of deep neural network (DNN) models. Several approaches have been proposed in recent years to adapt well-trained DNNs from a \textit{source} domain to a new \textit{target} domain, perhaps with limited training data.  Examples include data augmentation strategies \cite{kim17_far_field_generation}; the use of auxiliary features such as i-vectors \cite{peddinti15_reverb_ivectors}, posterior or bottleneck features \cite{thomas10_posterior_features,bell12_mlan} trained on {source}-domain data; adapting selected parameters \cite{yao12_adapt,swietojanski2016lhuc, samarakoon16_factorised}; adversarial methods \cite{shinohara16_adversarial}; as well as simple yet effective approaches such as applying further rounds of training to DNNs initialised on {source} data. 


The common ground in the vast majority of these works is that some transcribed data -- even if usually a limited amount --  from the {target} domain is available for adaptation of the acoustic models. This assumption, reasonable for well-resourced languages (WR), may not hold in the case of low-resourced languages (LR) for which even the amount of data available in the {source} domain may be very limited, and it is expensive or impractical to arrange for transcription of data from a new domain.

This is the scenario tackled in the IARPA MATERIAL programme\footnote{https://www.iarpa.gov/index.php/research-programs/material}.  The programme seeks to develop methods for searching speech and text in low-resource languages using English queries. In particular,  ASR systems must operate on diverse multi-genre data, including telephone conversations, news and topical broadcasts. However, 
the only manually annotated training data available is from the telephone conversations domain.

One approach to this problem is to collect a corpus of untranscribed data from the {target} domain in the LR language (for example, by web-crawling) and use an ASR system built for the {source} domain to create an automatic transcription, which is then used to train domain-adapted models.  This semi-supervised approach to DNN training has been successfully used eg. \cite{vesely13_semi_supervised, drugman16_semi_active,carmantini19_untranscribed}.  However, the technique requires careful confidence-based data selection, and is very sensitive to the performance of the {source} system on the {target} data.  Another drawback, when rapid deployment to a new domain is required, is the need to run computationally expensive decoding on large quantities of data in order to harvest sufficient quantities of training material. 

In this work, inspired by the challenges posed by the \mbox{MATERIAL} programme, we adopt a completely different approach: we explore whether it is possible to transfer a specific domain transform learned in a WR language to a LR language for which no {target} training data is available at all, in other words, is a method for adaptation between two given domains portable across languages?  We thus aim to improve the performance of a LR ASR system in a new \textit{target} domain by using only data of a WR language in both the \textit{source} and \textit{target} domains.
To this end, we propose an adaptation scheme that uses multi-lingual AM training to enable cross-lingual sharing of domain adaptation techniques.
Then, based on the hypothesis that initial layers of a DNN encode language-independent acoustic characteristics, we are able to transfer the adapted layers learned for one {target} domain from one language to another.

\begin{figure*}[t]
\centering{
\includegraphics[width=\linewidth]{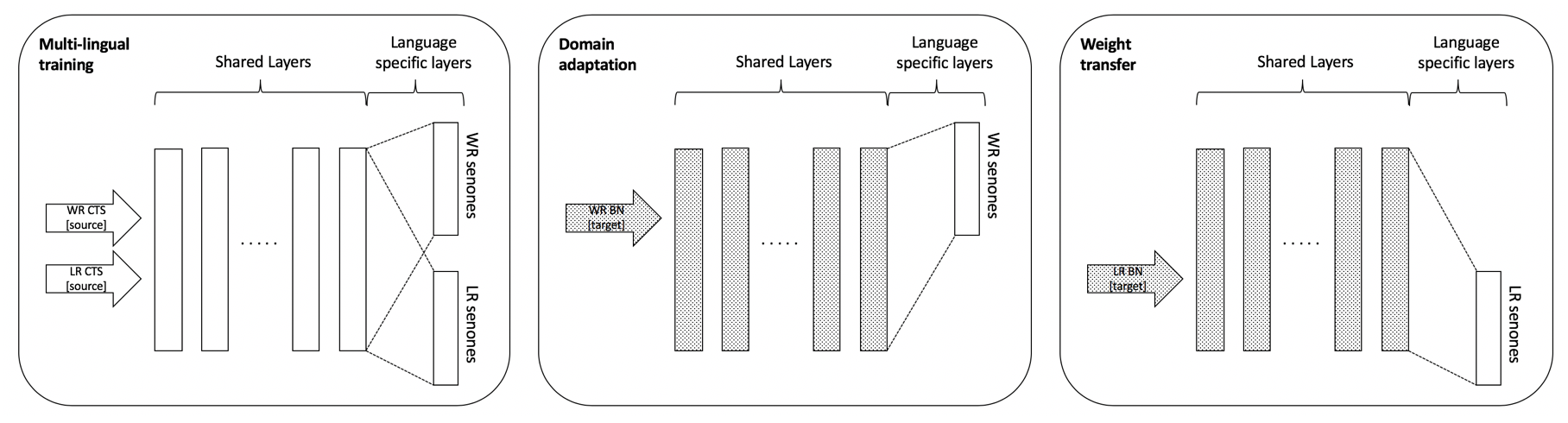}
\caption{Steps of the proposed cross-lingual domain adaptation scheme: 1) multi-lingual training; 2) adaptation of the shared parameters using WR data in the {target} domain; and 3) weight transfer of the domain adapted shared layers to the original LR final, language-dependent layers.}}
\label{fig:approach}
\end{figure*}

For the development of the proposed cross-lingual domain adaptation approach, it is more convenient to 
select a pair of languages for which data is available from both source and target domains in each language, enabling oracle
experiments to be carried out.
\added[id=aa]{Hence}, in this work we initially use English as the WR language
and pretend that Spanish is an LR language.  As in the real MATERIAL task, we choose conversational telephone speech (CTS) as the {source} domain and use broadcast news (BN) as the {target} domain.
We then explore the effectiveness of the proposed method on two of the actual MATERIAL target languages: Tagalog and Lithuanian.


The rest of this paper is organized as follows. Section \ref{sec:approach} describes the proposed cross-lingual domain adaptation approach. Then, experimental setup, including corpora and details on the architecture of the developed ASR systems, is reported in section \ref{sec:setup}. Finally, experimental evaluation is presented in section \ref{sec:results} before the final concluding remarks.

\section{Cross-lingual domain adaptation}
\label{sec:approach}

The main objective of this work is to propose an adaptation scheme for  DNN-based acoustic models that allows for cross-lingual domain adaptation from a LR language system trained for one \textit{source} domain into a \textit{target} domain using solely adaptation data of a WR language.
Considering that the role of the DNN is to learn a non-linear mapping between the acoustic features (e.g. MFCC) and phoneme-related classes (e.g. senones),
it is a common interpretation that the initial layers of a phonetic network are expected to encode lower-level acoustic information,  while deeper layers codify more complex cues closer to phonetic classes \cite{bengio13_representation}. 
Thus, following this interpretation,  we hypothesize that the initial layers of a phonetic network encode basic acoustic information  that is independent of the language and the task at hand, while the later ones are specific to each language. Under this interpretation, we suggest that the modifications that could be applied to the initial layers of a phonetic network in any language to adapt to the specific characteristics of a new domain should be similar (and transferable) among different languages.
To take advantage of this possibility, it is necessary to design a network architecture in which parameter transforms can be meaningfully shared among the LR and WR languages, followed by a set of final language specific layers.  This solution can be attained through multi-task learning and it is the backbone of our proposed scheme. As can be seen in Figure \ref{fig:approach}, the process consists mostly on three steps: 1) a multi-lingual network is trained with data of both the LR and WR language in the {source} domain (left); 2)  shared layers are adapted using WR data of the {target} domain (center); and 3) the adapted shared layers are transferred to the original LR language network resulting in a domain adapted version of the LR network (right).

\subsection{Multi-lingual training}

Multi-task learning \cite{caruana97_multitask} refers to the process of simultaneously learning multiple tasks from a single data set that contains annotations for different tasks. Typically, the network architecture consists of some initial layers that are shared by the multiple tasks and some final task-specific layers, one for each considered task. Back-propagation is applied for each task alternatively using all the training data propagated through both the task-specific and shared layers. This type of learning provides an improved regularization effect of the resulting networks compared to conventional single task learning approaches and has been used successfully in single-language acoustic modelling \cite{ seltzer2013multi, bell2017multitask}.

This type of approach has been also successfully applied in ASR using data from multiple languages to learn multi-lingual networks, in which each task objective corresponds to the phoneme (senone) classification of the different languages \cite{tuske13_multilingual, huang13_mutli_hybrid}.
In general, multi-lingual learning has shown to be particularly beneficial when languages with limited training resources are involved.
In this work, we use multi-lingual learning to train an initial network using data from both the LR and the WR languages in the {source} domain. Hence, the {source} multi-language network has a set of shared layers followed by two language specific (LR and WR) set of final layers. 


\subsection{Domain adaptation}


Given the multi-lingual architecture described previously, we adapt the shared components of the network using {target} data of the WR language whilst freezing the remaining language-specific layers. By doing so, our expectation is that the kind of transformations that the new adapted network will learn will be language-independent and will mostly be related with the particular acoustic characteristics of the new data.  Keeping the upper
layers frozen ensures that the newly adapted lower layers of the network will continue to be appropriate as inputs for the layers specific to the LR
language, despite no LR data being used for the adaptation.

In general, one may explore any of the well-researched strategies for network weight adaptation, such as LHUC \cite{swietojanski2016lhuc} or LIN \cite{li10_lin}.
In this work, given that substantial quantities of {target} domain data  are available in the WR language, we simply adapt the weights of a selected subset of layers of the shared network (initialized with the weights learned in the multi-lingual learning stage) through simple backpropagation updates with a varying number of data training epochs and an appropriately chosen learning rate.


\section{Experimental setup}
\label{sec:setup}


\subsection{Corpora}

In all experiments we take conversational telephone speech (CTS) as the {source} domain, for which transcribed training data
is available for all languages.  For English and Spanish, we train on data from the Fisher corpus\cite{cieri2004fisher} (\texttildelow200 hours and \texttildelow163 hours respectively).  Note that for the former, we use only a subset of the full corpus.  For Tagalog and Lithuanian, we use data from the IARPA Babel full language packs (80 hours and 40 hours respectively).
CTS data is all sampled at 8khz.  In the MATERIAL task, the target domain is a mixture of broadcast news (BN) and topical broadcast (TB) domain, both with wideband 16khz audio.  We approximate this target in English and Spanish by using broadcast news (BN) data from HUB4 \cite{graff19971996} with \texttildelow150 hours of English data, used for adaptation, and \texttildelow30 hours of Spanish, used for oracle experiments only.  For each corpus we use the standard evaluation sets: the 1997 HUB4 English Evaluation set is used for English BN; and for the Tagalog and Lithuanian, we use the BN and TB ``Analysis'' test sets provided by the MATERIAL programme.

\subsection{System description}

The Kaldi toolkit \cite{povey11_kaldi} has been used for the development of all the ASR systems.
To obtain the set of language-specific senones and frame-level phonetic alignments needed for training the DNNs, initial HMM-GMM systems have been built for each language and domain. HMM-GMM  training follows the conventional recipes in Kaldi, consisting on several stages of refinement from monophone to context-dependent models trained on  LDA+MLLT+fMLLR features \cite{Rath2013ImprovedFP}. 
\deleted[id=aa]{These  features are obtained  by splicing together 9 frames of 13-dimensional MFCCs, projecting down to 40 dimensions using linear discriminant analysis (LDA), normalizing the MFCCs to have zero mean per speaker, applying maximum likelihood linear transform (MLLT) and, finally, applying speaker adaptive training (SAT) using a single feature-space maximum likelihood linear regression (fMLLR) transform estimated per speaker. 
More details can be found in  \cite{Rath2013ImprovedFP}.}
\deleted[id=aa]{All t}
HMM-DNN ASR systems share a common input feature representation of 43 dimensions corresponding to 40 high resolution MFCCs components plus 3 additional pitch and voicing related features \cite{pitch_features_kaldi}. Neither side speaker information (i-vector), nor speed perturbation data augmentation have been used in these experiments. Note that all data is downsampled to 8 kHz to match the sampling rate of the CTS {source} domain data. Hence, all the systems reported in this work have been trained and evaluated at 8kHz.

The acoustic models are TDNN networks trained with frame-level cross-entropy loss criterion \cite{tdnn_2015}.
\deleted[id=aa]{The network architecture of the baseline single language systems consists of a first LDA transformation of the input (spliced with 2 left and right additional context frames), followed by 5 TDNN layers of 650 hidden units with RELU activations with different activation context configurations.}
\added[id=aa]{Network architectures consist}
\deleted[id=aa]{The  multi-lingual AM network, consists} 
of a stack of 7 TDNN hidden layers, each containing 650 units, with RELU activation functions. These correspond to the shared language-independent layers of the network. For each language, a pre-final fully connected layer of 650 units with RELU activations and a final softmax layer is appended. The size of the output layers correspond to the size of the outputs of the single language networks of the {source} CTS domain. During training, the samples of each language are not scaled, thus no compensation for different training data sizes is performed.

The optimization method used is natural gradient stabilized stochastic gradient descent \cite{povey2014parallel} with an exponentially decaying learning rate. The starting learning rate is set to 0.0015 and decays by a factor of 10 over the entire training. The baseline and multilingual AMs were trained for 3 epochs with a  minibatch size of 256. The parameter change is limited to a maximum of 2 for each minibatch to avoid parameter explosion. 
In the adaptation stage of the proposed approach, the network is initialized with the  multi-lingual network weights and the learning rate of the frozen layers is set to 0. The number of adaptation epochs is a varying parameter investigated in the results section.
All the remaining configurations are identical to that of the multi-lingual training stage. 

Since the focus of this work is on adaptation of the AM, domain matched language models are always used in decoding. That is, CTS {source} and BN {target} domain test data is decoded with corresponding language-specific CTS and BN LMs. 
While this seems to go against the assumption of working with low resource languages, text data is usually much easier to obtain compared to transcribed speech.  
CTS LMs are trained on the training transcriptions of the relevant corpora. BN LM for Spanish is also trained only on the training transcriptions, while the BN LM for English is trained using the transcriptions and additional text from the 1996 CSR HUB-4 Language Model and the North American News Text Corpus.  
BN LMs for Lithuanian and Tagalog are trained on around 30M words of web-crawled text from CommonCrawl and other online sources.




\section{Results}
\label{sec:results}

\subsection{English and Spanish baseline systems}

In this and the following two sections, we take Spanish to be the LR language; as always, the WR language is English.
Table \ref{tab:multilang} shows, in the {first} row, word error rate (WER) performance of  single language baseline systems using BN {target} in-domain acoustic models.
In the case of the LR language, this is an oracle experiment, since we assume that in reality, no {target} domain data is available for this language. 
The {second} row shows results of  
CTS {source}  domain acoustic models evaluated with both CTS and BN test data. 
As expected, one observes a large degradation when decoding LR BN {target} domain data with CTS acoustic models: about 20\% absolute WER compared with the matched experiments. 
The performance drop can also be  observed in the WR experiments, but it is not so acute, probably due to the increased amount of training data used in the WR system. 
Note that the objective of this work is to make the 40.0\% of the LR baseline system when decoding {target} domain data as close as possible to the oracle 19.2\% figure, without using any in-domain LR language training data.

\begin{table}[t]
\centering
\begin{tabular}{l|C{0.10\linewidth}|C{0.10\linewidth}||C{0.10\linewidth}|C{0.10\linewidth}|}
\cline{2-5}
 & \multicolumn{4}{c|}{Test condition} \\ \cline{2-5} 
  & \multicolumn{2}{c||}{WR} & \multicolumn{2}{c|}{LR} \\ \cline{2-5} 
  & CTS {source} & BN {target} & CTS {source}  & BN  {target}    \\ \hline
\multicolumn{1}{|l|}{mono-ling BN AM} & --- & 11.8 & --- & 19.2 \\ \hline
\multicolumn{1}{|l|}{mono-ling CTS AM}  & 22.6  & 19.6 & 32.3  & 40.0 \\ \hline
\multicolumn{1}{|l|}{multi-ling CTS AM} & 23.6  & 19.2  & 32.6  & 32.9  \\ \hline
\end{tabular}
\caption{\added[id==a]{WER (\%) of the mono-lingual matched BN target domain trained systems (first row), the mono-lingual mismatched CTS source domain trained systems (second row) and the multi-lingual ASR system trained with WR and LR CTS {source} data (third row) for the different test domain conditions.}}
\label{tab:multilang}
\end{table}


\added[id=aa]{The last row of} Table \ref{tab:multilang} shows the WER performance of the multi-lingual system trained with LR and WR data of the CTS {source} domain.
\deleted[id=aa]{(second row) in contrast to the single language baselines reported previously (first row)}
For both languages, the performance of the multi-lingual models on the CTS source domain is close to that obtained using the respective mono-lingual model. 
\deleted[id=aa]{: slight decrease in the WR case (0.5\% performance drop) and slight increase in the LR case (0.5\% performance improvement).} 
However, for the cross-domain test case, there is a remarkable improvement in the LR language performance from 40.0\% to 32.9\%.
This is a 17.8\% relative improvement in the BN {target} domain achieved by using only out-of-domain CTS WR data, thanks to the multi-lingual training scheme.
While the benefits of multi-lingual training for LR were expected,  it is very interesting to observe that these are much more significant in the cross-domain case. This may be partially explained due to the considerably increased amount of data and variation to which this network is exposed compared to the LR baseline.

\subsection{Cross-lingual network adaptation results}

\deleted[id=aa]{Table \ref{tab:proposed} shows WER performance in the LR BN {target} test set applying}
\deleted[id=aa]{Recall that this approach solely uses WR BN {target} domain data for adaptation of the network. 
For each combination of number of epochs and layers, the performance in the WR BN {target} test set is also reported in parenthesis.}
\added[id=aa]{The third row of Table \ref{tab:comparison} reports the  performance of the proposed method in contrast to the mono-lingual (first row) and multi-lingual (second row) baselines.}  \added[id=aa]{The proposed cross-lingual domain adaptation method has been tested for} 
a varying number of training epochs (from 0.5 to 3) and adapted shared layers (from 1 to 6); the reported result corresponds to the best adaptation configuration, which is obtained when the 3 first hidden shared layers are adapted using all WR {target} domain data for 1 epoch of training. We observed
in the complete set of experiments that
results on LR data are not particularly sensitive to number of epochs or layers amongst those tested, ranging from 28.4\% to 29.1\% in all cases. However, as expected, this is not the case for the WR language results, in which performance tends to keep improving with increased number of adapted layers and epochs. Hence, an improvement in the WR case does not necessarily imply an improvement in the LR case.
We conclude that there seems to be a limit on the amount of information that is transferable from the WR system to the LR system. 
\deleted[id=aa]{Overall, the best results on the LR BN {target} test set are obtained when the 3 first hidden shared layers are adapted using all WR {target} domain data for 1 epoch of training.}
\deleted[id=aa]{Finally, for ease of comparison, we summarize the main results attained}
\added[id=aa]{Overall, we observe that by}
using WR CTS {source} domain data for multi-lingual learning we are able to improve from 40.0\% to 32.9\% and by then using WR BN {target} domain data and the proposed adaptation method, we further increase performance to 28.4\% WER. This is an absolute 11.6\% WER decrease, which is a recovery of around 50\% of the performance loss due to the lack of LR training data in the CTS {target} domain when compared to a system fully trained with BN data (see Table \ref{tab:multilang}). 
This is attained by using only WR data and no any additional LR data.

\begin{table}[t]
\centering
\resizebox{0.9\columnwidth}{!}{%
\begin{tabular}{l|C{0.2\linewidth}|C{0.2\linewidth}|}
\cline{2-3}
  & WR   & LR   \\ \cline{2-3} 
  & BN  {target}   & BN  {target}  \\ \hline
\multicolumn{1}{|l|}{mono-ling CTS AM (1)} & 19.6 & 40.0 \\ \hline
\multicolumn{1}{|l|}{multi-ling CTS AM (2)}  & 19.2 & 32.9\\ \hline
\multicolumn{1}{|l|}{proposed CL adapt AM (3)} & {14.5}  & \textbf{28.4} \\ \hline
\multicolumn{1}{|l|}{multi-task CL AM (4)} & {12.4}  & {29.1} \\ \hline
\multicolumn{1}{|l|}{multi-task CL + adapt AM (5)} & {12.3}  & {29.1} \\ \hline
\multicolumn{1}{|l|}{multi-cond CL AM (6)} & {12.5}  & {29.2} \\ \hline
\multicolumn{1}{|l|}{multi-cond CL + adapt AM (7)} & {12.2}  & {29.1} \\ \hline
\end{tabular}
}
\vspace{-2mm}
\caption{\added[id=aa]{WER (\%) of the mono-lingual  AM (1), the multi-lingual AM (2), the proposed adapted AM (3) and alternative cross-lingual  AM (4-7) obtained in the WR and LR BN {target} domain test  sets.}}
\label{tab:comparison}
\vspace{-1mm}
\end{table}

\subsection{Comparison with similar cross-lingual approaches}

\added[id=aa]{In this section, the proposed approach is compared with two related cross-lingual information transfer methods:
first, with an AM trained in a multi-task fashion, considering  LR source, WR source and WR target as three separate tasks, referred to as \emph{multi-task}; and second, with an AM trained in a multi-task and multi-condition fashion, considering the LR source as one task, and the mix of WR source and WR target as a second task, referred to here as \emph{multi-cond}. 
Rows 4 and 6 in
Table \ref{tab:comparison} report  performance  of these two cross-lingual alternative approaches.
For these experiments, the exact same network architecture, training and decoding recipes as previously have been followed. 
Notice that, like in the proposed method, it is  possible to use these networks as an initialization for further fine-tuning  using only WR target domain data for a varying number of epochs and shared adaptation layers. Thus, rows 5 and 7 of Table \ref{tab:comparison} report results after fine-tuning adaptation for the best configuration of epochs and number of adapted layers in each case. 
For both approaches, the additional fine-tuning does not provide significant improvements. Performance converges already after the initial training.  In fact, we observe that the best adaptation configuration is attained with the minimal number of epochs and adaptation layers. For any other configuration, performance oscillates in absolute differences of $\pm0.1$.
Overall, the cross-lingual proposed scheme outperforms any of the other methods, being able to better leverage  information from the WR data for improved  LR acoustic modelling in the target domain.
}

\subsection{Experiments with MATERIAL languages}
\added[id=aa]{
In this section we investigate the proposed cross-lingual adaptation approach considering two of the IARPA MATERIAL languages: Tagalog and Lithuanian. 
For these experiments, we keep the same network architecture and training and decoding recipes as previously, including the set of best parameters found for the proposed cross-lingual method: adaptation of the 3 first hidden shared layers  for 1 epoch.
The new languages are less related to English than Spanish, and the data available present a poorer match between source and target conditions. 
Despite the significant differences among languages and target domain conditions, 
results reported in Table \ref{table:material} show that the proposed method is able to effectively exploit English data to improve ASR performance of LR languages in any of the wideband data sub-conditions. 
As expected, the proposed method is more effective in closer target conditions to those of the WR data: for the BN wideband sub-condition the relative WER improvements are 21.2\% for Tagalog and 30.7\% for Lithuanian; while the  improvements for the TB wideband sub-condition  are 17.4\% for Tagalog and 25.3\% for Lithuanian.  
Overall, the average relative WER improvements for the wideband conditions are 18.3\% and 27.5\% for the Tagalog and Lithuanian languages, respectively.
}

\begin{table}[t]
\resizebox{\columnwidth}{!}{%
\begin{tabular}{l|c|c|c|c|c|c|}
\cline{2-7}
 & \multicolumn{3}{c|}{Tagalog} & \multicolumn{3}{c|}{Lithuanian} \\ \cline{2-7} 
 & BN & TB & Avg. & BN & TB & Avg. \\ \hline
\multicolumn{1}{|l|}{mono-ling CTS (1)} & 53.2 & 58.7 & 57.3 & 45.6 & 43.0 & 44.0 \\ \hline
\multicolumn{1}{|l|}{multi-ling CTS (2)} & 46.5 & 52.2 & 50.7 & 38.2 & 36.5 & 37.1 \\ \hline
\multicolumn{1}{|l|}{proposed CL adapt (3)} & 41.9 & 48.5 & \textbf{46.8} & 31.6  & 32.1 & \textbf{31.9} \\ \hline
\end{tabular}
}
\vspace{-2mm}
\caption{\added[id=aa]{WER (\%) of the single language  AM (1), the multi-lingual AM (2) and the proposed adapted AM  in the Tagalog and Lithuanian MATERIAL evaluation sets.}\label{table:material}}
\vspace{-1mm}
\end{table}


\deleted[id=aa]{
In spite of the encouraging results reported in this section, there are some remaining open questions that still need to be explored. For instance, the proposed method needs to be compared with other cross-lingual information transfer methods, such as bottle-neck features trained on multi-lingual multi-domain data. Note that our method can be interpreted as a sort of adaptive posterior-based feature extractor.
Similarly, SAT vector-based approaches (e.g. i-vectors) may indirectly contribute to some domain compensation that needs to be assessed in combination with our approach.
Second, the adaptation approach needs to be evaluated with current state of the art sequence training methods, such as LF-MMI \cite{povey_lfmmi}. Nevertheless, as there is evidence proving the validity of information transfer and multi-task learning for those techniques \cite{ghahremani2017investigation}, we do not foresee any obstacles to the application of the proposed method. 
Finally, it might be also necessary to investigate the impact of the closeness of the two selected languages, since Spanish and English are relatively similar, and the similarity of the conditions on both languages in the {source} and {target} domains. Regarding the latter, a promising interesting future direction may consist of the combination of the proposed method together with approaches to cross-lingual unsupervised domain identification.
}
\section{Conclusions}
\label{sec:conclusions}

This paper has demonstrated that it is possible to transfer domain adaptation of DNNs from one language to another, enabling adaption of a low-resourced language to be performed with absolutely no data from the {target} domain.
This has been achieved thanks to a multi-lingual network architecture that allows for meaningful share of the parameter transforms among languages. In our experiments, the proposed cross-lingual domain adaptation approach \added[id=aa]{outperforms other similar methods achieving up to a 29\%} relative WER improvement in the {target} domain 
\added[id=aa]{when similar languages and source and target domain conditions are considered. Moreover, the proposed adaptation scheme also allows for remarkable WER improvements in the case of less favorable language and domain conditions. 
} 
Future work will extend the method to sequence-trained models and also investigate 
\added[id=aa]{the combination with other cross-lingual information transfer methods, such as bottle-neck features trained on multi-lingual multi-domain data, and SAT vector-based approaches (e.g. i-vectors).}
\deleted[id=aa]{whether the method can be extended to situations when matching domains across languages cannot be precisely identified.}

\vfill\pagebreak



\bibliographystyle{IEEEbib}
\bibliography{refs}

\end{document}